\documentclass[twocolumn]{aastex631}
\usepackage{amsmath}
\usepackage[version=3]{mhchem} 
\UseRawInputEncoding

\begin{document}

\title{Characterizing the Radiative-Convective Structure of Dense Rocky Planet Atmospheres}

\author[0009-0003-5970-570X]{Jessica Cmiel}
\affiliation{Harvard Paulson School of Engineering and Applied Sciences \\
29 Oxford Street\\
Cambridge, MA 02138, USA}

\author[0000-0003-1127-8334]{Robin Wordsworth}
\affiliation{Harvard Paulson School of Engineering and Applied Sciences \\
29 Oxford Street\\
Cambridge, MA 02138, USA}
\affiliation{Department of Earth and Planetary Sciences, Harvard University\\
20 Oxford Street\\
Cambridge, MA 02138, USA}

\author[0000-0003-0769-292X]{Jacob T. Seeley}
\affiliation{Department of Earth and Planetary Sciences, Harvard University\\
20 Oxford Street\\
Cambridge, MA 02138, USA}


\begin{abstract}

We use a one-dimensional line-by-line radiative-convective model to simulate hot, dense terrestrial-planet atmospheres. We find that strong shortwave absorption by \ce{H2O} and \ce{CO2} inhibits near-surface convection, reducing surface temperatures by up to $\sim\!2000$ K compared to fully convective predictions.
Pure \ce{CO2} atmospheres are typically 1000~K cooler than pure \ce{H2O}, with only a few percent of \ce{H2O} required to elevate surface temperatures by hundreds of kelvin for a fixed incident stellar radiation. 
We also show that minor greenhouse gases such as \ce{SO2} and \ce{NH3} have a limited warming effect when \ce{H2O} is abundant. 
We find that even for insolation values as high as 12500~W\,m$^{-2}$ (37 $\times$ Earth's), planets with mixed \ce{CO2}-\ce{H2O} envelopes have surface temperatures in the 1200\text{---}2000\,K range,  limiting surface melting. 
Our results highlight the critical role of shortwave heating on magma ocean planets and the need for improved high-temperature spectroscopy beyond 20{,}000\,cm\(^{-1}\).

\end{abstract}


\section{Introduction}
\label{sec:intro}

Radiative and convective processes are fundamental to energy transfer in planetary atmospheres. Radiative processes govern both incoming stellar radiation and outgoing thermal emission, ultimately setting atmospheric and surface temperatures for a given outgoing longwave radiation (OLR) \citep{Pierrehumbert2010}. Convection, in turn, transports heat vertically where radiative cooling or heating alone cannot maintain stability. On Earth, this vertical redistribution by convection dominates the lower atmosphere, helping shape circulation patterns and climate \citep{emanuel1988}.

The interaction of radiation and convection creates the layered structure characteristic of many terrestrial atmospheres, with radiative regions atop convective ones. Understanding these processes is key to predicting surface conditions, atmospheric structure, and global energy transport---particularly for exoplanets whose atmospheric compositions may diverge significantly from Earth's. Accurate modeling of these processes often requires non-gray radiative-convective models, which capture gas absorption over numerous spectral bands and thus handle the broad range of temperatures and pressures in exoplanetary environments \citep{Seager2010}.

Above a critical absorbed stellar flux, classical, simplified models of Earth-like planets have invoked a ``runaway greenhouse" transition---a threshold where water vapor becomes the dominant atmospheric constituent, and strong infrared opacity greatly amplifies surface warming \citep{Kasting1988}. The canonical runaway occurs once the atmosphere reaches an optically thick regime: increased stellar flux no longer boosts the OLR but instead drives further surface heating until all surface water evaporates. This hot, steam-dominated state is often referred to as the ``post-runaway" regime. In such scenarios, it has long been assumed that the atmosphere remains fully convective, producing surface temperatures sufficiently high to sustain large-scale magma oceans for extended periods \citep{lebrun2013,elkinstanton2012}.
However, recent work has challenged the assumption that post-runaway steam atmospheres are fully convective. Even below the classical runaway threshold, strong shortwave absorption by \ce{H2O} --- particularly in the near-infrared --- can form stable radiative layers that suppress convection near the surface \citep{Selsis2023cool}. By reducing the vertical temperature gradient, this radiative inhibition decouples the surface from convective cooling, limiting the maximum achievable surface temperature at a given OLR. If the surface remains below the mantle solidus, long-lived global magma oceans may be curtailed or entirely prevented; conversely, if shortwave absorption keeps the surface well above the solidus, it can prolong the magma-ocean phase, influencing planetary outgassing, volatile retention, and interior cooling \citep{Abe1985, ElkinsTanton2008, hamano2013emergence, lebrun2013, Nikolaou2019, Bower2021, Turbet2021, Kite2021}.

This suppression of convection arises when intense near-infrared absorption warms upper layers enough to reduce or even reverse the usual temperature lapse rate, preventing buoyant updrafts. While this effect is well documented for pure-steam atmospheres, realistic planetary envelopes often contain multiple greenhouse species. In particular, \ce{CO2}, a dominant volcanic outgassing product, can accumulate on rocky planets where hydrogen escape has oxidized the surface or carbon sinks are inefficient \citep{Seager2010, Hu2014,sossi2020redox}. It is therefore crucial to determine whether mixed \ce{H2O}-\ce{CO2} atmospheres exhibit similar radiative inhibition and how even small \ce{H2O} fractions might restore near-infrared opacity to reshape surface conditions. 

Multiple processes can supply \ce{H2O} to planetary atmospheres, including exsolved volatiles from magma oceans, serpentinization reactions, exogenous delivery by impacts, and high-temperature reactions between Fe and \ce{H2} in reducing conditions \citep[e.g.,][]{Schaefer2007, Hirschmann2012, GaillardScaillet2014}, while \ce{CO2} is a primary volcanic outgassing product, particularly in oxidizing conditions \citep{Hu2014,sossi2020redox}.

Over time, hydrogen escape can drive atmospheric oxidation, leading to \ce{O2}- or \ce{CO2}-rich compositions \citep{Wordsworth2018}. Finally, trace gases such as \ce{SO2} or \ce{NH3} released by volcanism may also influence absorption in key wavelength ranges \citep{Gao2017}. These geochemical processes shape the long-term evolution of planetary atmospheres, particularly in the transition from steam-rich to oxidized states.

In this paper, we focus on dense atmospheres with surface temperatures exceeding \(\sim\)1000~K, a regime relevant to highly irradiated planets and early evolutionary phases.
We use a high-resolution, line-by-line (LBL) radiative-convective model to revisit and extend earlier work \citep{Selsis2023cool}.

Critically, we aim to clarify whether the decoupling of surface and atmosphere inferred for pure-steam envelopes carries over to more realistic compositions, how it affects magma ocean existence or volume, and how minor species factor into the energy budget when water's near-infrared bands dominate. We then discuss the implications for interior cooling, volatile outgassing, and potential climate ``hysteresis" or multi-stability in high-temperature regimes.

Our paper is organized as follows. Section~\ref{sec:methods} describes our model, including updates for high-temperature line-by-line calculations and hybrid vertical discretization. Section~\ref{sec:results} presents the resulting temperature-pressure profiles for \ce{H2O}, \ce{CO2}, and mixed atmospheres, highlighting how shortwave absorption suppresses near-surface convection. In Section~\ref{sec:discussion}, we address how this affects magma-ocean longevity, the total volume of molten material, and the broader planetary evolution perspective. Finally, we conclude in Section~\ref{sec:conclusion} by summarizing key insights and identifying directions for future work.

\begin{figure}[ht!]
    \centering
    \includegraphics[width=\columnwidth]{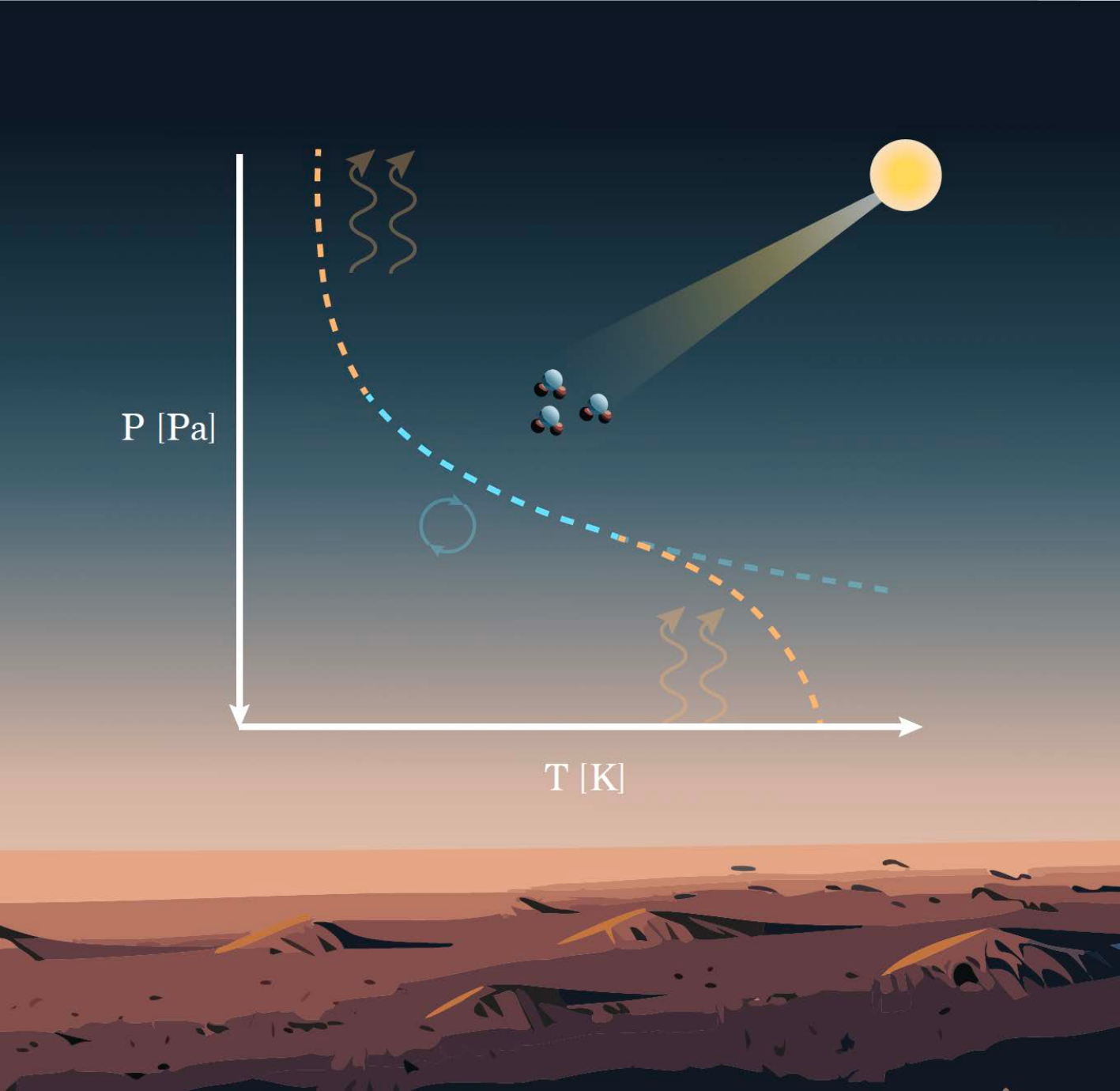}
    \caption{Schematic depiction of an atmospheric temperature profile (dotted line) showing convective regions (blue) and radiative regions (orange). The profile illustrates a detached convective region in the middle atmosphere, separated from the surface by a radiative layer.}
    \label{fig:schematic}
\end{figure}

\section{Methods}
\label{sec:methods}

\subsection{Model Overview}
Our goal is to capture how radiative and convective energy transfer shape the thermal structure of dense, hot exoplanet atmospheres. We adopt a high-resolution, line-by-line (LBL) radiative transfer approach to resolve individual absorption lines and continuum features, which become increasingly important as temperature and pressure rise. While correlated-$k$ methods can be computationally faster, a LBL approach is more flexible and accurate and allows the latest linelists for high-temperature atmospheres to be implemented and rapidly tested \citep{Wordsworth2021, Ding2019}.

We implement our model in a version of PCM-LBL, which we call PCM-HiPT (High-Pressure/Temperature). The core upgrades include:
\begin{enumerate}
    \item An iterative convection scheme operating pairwise between adjacent vertical layers (Section~\ref{subsec:conv_scheme}) designed for large temperature gradients and possible detached convective zones.
    \item A hybrid pressure grid for fine resolution near the surface (Appendix~\ref{appendix}), where high optical depths demand smaller layer spacing.
    \item A refined flux convergence method to maintain a fixed outgoing longwave radiation (OLR), which isolates how different atmospheric compositions affect near-surface temperatures.
\end{enumerate}

This setup is motivated by the extreme regimes we explore: surface temperatures often exceed 1000~K, requiring careful handling of high-pressure absorption lines. Our overarching aim is to understand whether strong shortwave absorption by \ce{H2O} (and possibly \ce{CO2}) can suppress near-surface convection and thereby affect magma ocean stability.

\subsection{Convection Scheme}
\label{subsec:conv_scheme}
A critical question for hot exoplanets is whether convection remains coupled to the surface or is ``detached" by intense shortwave heating in upper layers. To capture this possibility, we replace the traditional approach of forcing all unstable layers onto a single adiabat \citep[e.g.,][]{Manabe1964} with an iterative, pairwise convective adjustment scheme.

Following \citet{Ding2016}, we calculate the potential temperature in each layer and identify where it decreases with altitude, indicating instability. We then adjust only those adjacent layers to remove the instability, rather than forcing the entire column to share a single lapse rate. This local approach allows multiple separate convective regions, one of which can be ``detached" from the surface if shortwave absorption creates a stable layer below.

Specifically, we update the temperatures of layers $i$ and $i+1$ (denoted $T_i'$ and $T_{i+1}'$) according to:
\begin{equation}
    T_i' = \theta' \left(\frac{p_i}{p_0}\right)^{\tfrac{R_s}{c_p}}, 
    \quad
    T_{i+1}' = \theta' \left(\frac{p_{i+1}}{p_0}\right)^{\tfrac{R_s}{c_p}},
\end{equation}
and
\begin{equation}
    \theta' \;=\; \frac{
    T_i \Bigl(\tfrac{p_0}{p_i}\Bigr)^{\tfrac{R_s}{c_p}}\,\Delta p_i 
    \;+\;
    T_{i+1}\Bigl(\tfrac{p_0}{p_{i+1}}\Bigr)^{\tfrac{R_s}{c_p}}\,\Delta p_{i+1}
    }
    {\Delta p_i + \Delta p_{i+1}}.
\end{equation}
Here, $\Delta p_i$ is the pressure thickness of layer $i$, $R_s$ is the specific gas constant, and $c_p$ is the (constant) heat capacity. We iterate through the column until no further instabilities remain, thus conserving energy while accommodating multiple convective zones.

We set $p_0$, the ``reference pressure" used in potential temperature computations, to the surface pressure in each simulation. This choice ensures that the pairwise convective adjustment conserves energy layer by layer and keeps calculations self-consistent for the entire atmospheric column. Sensitivity checks indicate that alternate definitions (e.g., a fixed 1~bar reference) does not change the results.

We assume a compositionally well-mixed atmosphere. We compared our temperature profiles with the saturation vapor pressure curves of \ce{H2O} and \ce{CO2} to check that condensation did not occur in our simulations.  For \ce{H2O}, the saturation vapor pressure follows the parameterization of \citet{MurphyKoop2005}, which provides an empirically validated expression applicable across a broad range of atmospheric temperatures.  For \ce{CO2}, the condensation temperature as a function of pressure follows an empirical formulation. Below the triple point pressure (518,000 Pa), we adopt the sublimation relation from \citet{Fanale1982}, while at higher pressures, we use a liquid-vapor transition model based on data from the CRC Handbook \citep{CRCHandbook2003}. These expressions allow us to accurately capture phase transitions for \ce{CO2} over the full range of pressures relevant to our simulations.
As shown later (e.g., Fig.~\ref{fig:atmosprofco2}), the modeled temperature profiles did not intersect these condensation curves, so we neglected latent heating associated with condensation. 
\subsection{Pressure Scheme}
Dense atmospheres with large vertical pressure ranges pose a numerical challenge: one must resolve strong gradients near the surface while avoiding an unwieldy number of layers aloft. To address this, we employ a ``hybrid" pressure grid that merges linear spacing (in the high-pressure region) with logarithmic spacing (at lower pressures).

In practice, we define a transition pressure level and switch smoothly from linear increments (fine resolution where optical depth is highest) to log increments (coarser resolution where the atmosphere is thinner).Appendix~\ref{appendix} details the weighting functions and example arrays. This ensures we capture near-surface heating accurately while maintaining computational efficiency.

\subsection{Radiation Scheme}
PCM-HiPT employs a line-by-line method to calculate absorption over a dense wavenumber grid, reflecting updated spectroscopic data from HITRAN2020 \citep{Gordon2022}. We generate look-up tables for \ce{H2O} and \ce{CO2}, then interpolate them at each atmospheric layer's temperature and pressure. 

\subsection{Model Validation with a Gray Gas Benchmark}
\label{subsec:gray_validation}

Before applying our model to real-gas compositions, we first test it in a simplified, analytically tractable case: a gray gas atmosphere with shortwave absorption. This serves as a benchmark to confirm that PCM-HiPT reproduces established radiative-convective behavior, ensuring that numerical solutions align with theoretical expectations.

Following \citet{rc12}, we consider a gray gas with an imposed longwave-to-shortwave opacity ratio of $\kappa_{\mathrm{LW}} / \kappa_{\mathrm{SW}} = 10$. This setup approximates an idealized planetary atmosphere where shortwave absorption is moderate but does not fully dominate, making it a useful reference case for understanding how near-surface radiative layers form. The qualitative behavior remains consistent across a wide range of opacity ratios. Tests spanning \(\kappa_{\mathrm{LW}} / \kappa_{\mathrm{SW}} = 10^{-3}\) to \(10^2\) confirm that while the altitude of the radiative layer shifts depending on the shortwave absorption strength, the fundamental presence of radiative inhibition persists.

\begin{figure}[ht!]
    \centering
    \includegraphics[width=\columnwidth]{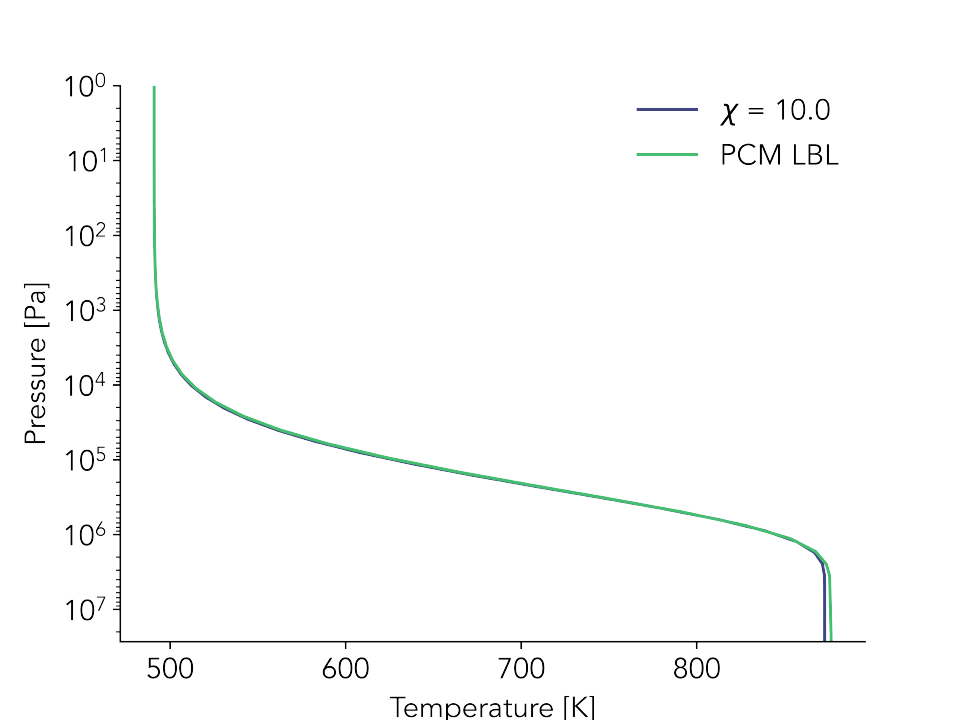}
    \caption{Comparison of radiative temperature profiles for a gray gas with $\kappa_\text{LW}/\kappa_\text{SW} = 10$. The figure shows curves generated using PCM-HiPT (solid line) and an analytic gray gas model (dashed line) with shortwave absorption. The close agreement shows that our radiative-convective model correctly handles gray atmospheres where shortwave absorption modifies convection.}
    \label{fig:graygascomparison}
\end{figure}

Figure~\ref{fig:graygascomparison} compares PCM-HiPT's temperature-pressure profile to an analytic gray gas model with shortwave absorption. The results show close agreement, confirming that the radiative-convective scheme correctly balances energy fluxes and captures the expected formation of stable radiative regions above the convective zone.

This validation demonstrates that our model accurately reproduces the fundamental physics of radiative-convective equilibrium, including cases where strong shortwave absorption modifies the convective structure. It provides confidence in applying PCM-HiPT to more complex, real-gas atmospheres with mixed \ce{H2O}--\ce{CO2} compositions.

\subsection{Equilibration Method}
We fix the OLR at 10,000\,W\,m$^{-2}$, iterating to find an incident stellar radiation (ISR) that ensures net radiative equilibrium at the top of the atmosphere (ASR = OLR). This approach isolates compositional effects by holding OLR constant across different scenarios. After each iteration, we compare the absorbed solar radiation (ASR) to the target OLR, updating the ISR by a factor $\beta$:
\begin{equation}
\beta \;=\; 1 + \frac{\bigl(\mathrm{OLR} - \mathrm{ASR}\bigr)\,\Delta t}{\alpha\,\mathrm{ISR}},
\end{equation}
where $\Delta t$ is the timestep, $\alpha$ is a damping coefficient, and $\mathrm{ISR}$ is the previous incident flux. We then rescale the downward stellar flux at each level:
\begin{equation}
I_{\text{lev, dn}}(\mathrm{lev}) \;=\; I_{\text{stel, dn}}\,\times\,\beta.
\end{equation}

This iterative procedure ensures that no net energy is artificially gained or lost and that the final temperature-pressure profile reflects the chosen OLR while satisfying both layer-by-layer radiative balance and overall equilibrium at the top of the atmosphere.

We deem the solution converged once the following three conditions are met: (1) the surface temperature remains stable within 0.1\,K over the final 10\% of timesteps; (2) OLR and ASR match to within 1\,W\,m$^{-2}$; and (3) OLR and ASR vary by no more than 0.5\,W\,m$^{-2}$ in the last 10\% of steps.

Since the bond albedo directly controls the absorbed solar radiation (ASR), its variation with atmospheric composition necessitates an adjustment in the incident stellar radiation (ISR) to maintain the prescribed radiative equilibrium (OLR = ASR). In cases where an increased bond albedo reflects more of the incoming flux (e.g. \ce{CO2}-dominated), a higher ISR is required to achieve the target ASR, and conversely, a lower ISR is needed when the bond albedo is reduced. 
Keeping OLR fixed allows us to compare varying composition atmospheres with the same absorbed energy, regardless of albedo. Conversely, keeping ISR fixed is most appropriate for investigating how radiative structure depends on composition for a given planet.

Here we mainly compare profiles with fixed OLR of $10^4\,\mathrm{W\,m}^{-2}$ except for in two cases (Figs. \ref{fig:atmosprofco2} and \ref{fig:melt_fraction}), where ISR was fixed at $10^4\,\mathrm{W\,m}^{-2}$.

\subsection{Water Vapor Continuum}
Continuum absorption, especially from water vapor, can significantly affect radiative transfer in hot, dense regimes. We use the MT\_CKD (Mlawer\-Tobin\-Clough\-Kneizys\-Davies) water vapor continuum model in our simulations  \citep{mlawer2023}. This semi-empirical model provides absorption coefficients for both self and foreign water vapor continua across a wide spectral range (0-20,000 cm$^{-1}$). We chose MT\_CKD for its comprehensive coverage of continuum effects, including the ``pedestal" effect at $\pm$25 cm$^{-1}$ from line centers. However, MT\_CKD has specific limitations: it may underestimate absorption in near-infrared windows and overestimate it in the far-infrared \citep{Paynter2011}. These inaccuracies could affect our calculations of radiative forcing, particularly in spectral regions critical for atmospheric energy balance. Additionally, the model's semi-empirical nature means it does not fully capture temperature and pressure dependencies in hot, dense atmospheres. These limitations are intrinsic to the MT\_CKD model and not specific to our implementation. 
We discuss in Section~\ref{sec:discussion} how these uncertainties could influence our high-temperature results, particularly for  hot surfaces and strong near-IR stellar fluxes.

\subsection{Heat Capacity}
\label{sec:cp_sensitivity}
We adopt a constant heat capacity, $c_p=2085$~J\,kg$^{-1}$\,K$^{-1}$, representative of high-temperature steam \citep{Haar1972}. In principle, $c_p$ can vary with temperature and pressure, but tests (Appendix~\ref{appendix:cp_sensitivity}) show that such variations shift the temperature profile by only a few tens of kelvin and do not remove the near-surface radiative zone. Thus, a fixed $c_p$ provides a sufficiently accurate baseline for our high-T scenarios.

\subsection{Rayleigh Scattering}

Rayleigh scattering is applied uniformly at the top of the atmosphere using an analytic conservative two-stream formula \citep{Pierrehumbert2011,wordsworth2017transient}. 
The total optical depth is derived by summing individual gas contributions weighted by their molar concentration and mixing ratio. Refractive indices are calculated following formulations by \citet{bideaumehu1973} for \ce{CO2} and \citet{huffman1971} for \ce{H2O}.

This scattering scheme can alter the shortwave radiation budget by increasing the planetary albedo in the blue-to-visible wavelength range, thereby reducing the flux that penetrates to the lower atmosphere. We find that in \ce{CO2}-rich atmospheres exposed to relatively high flux in the 0.4-0.7 $\mu m$ region (e.g., Sun-like or F-type stars), Rayleigh scattering modestly elevates the top-of-atmosphere albedo and can shift some fraction of the absorbed flux to higher altitudes. In turn, this slight redistribution of shortwave heating can affect the onset and thickness of convective layers. Our calculated bond albedo values show close agreement with \ce{CO2}-dominated cases reported by \citet{pluriel2019} (A = 0.46 versus their 0.45) , while we note a tendency to slightly overestimate albedo in \ce{H2O}-dominated scenarios (A = 0.30 versus their 0.25). For cooler stars with more near-infrared emission, scattering becomes less impactful because molecular absorption bands typically dominate over Rayleigh scattering, resulting in deeper flux penetration and enhanced near-surface heating. Consequently, the interplay of scattering and absorption determines whether a stable radiative layer forms immediately above the surface or deeper in the atmosphere.

\section{Results}
\label{sec:results}

\subsection{Pure \ce{H2O} Atmospheres: Comparison with Previous Studies}
\label{subsec:results_pure_h2o}

We begin by examining atmospheres composed entirely of \ce{H2O}, extending earlier findings that shortwave absorption can inhibit near-surface convection. These ``steam" atmospheres have historically been considered fully convective in post-runaway greenhouse scenarios \citep{Kasting1988}, but recent work \citep{Selsis2023cool} shows that strong near-infrared \ce{H2O} absorption can maintain a radiative layer near the surface, reducing $T_s$ relative to a purely adiabatic profile.

\begin{figure*}[ht!]
    \centering
    \includegraphics[width=\textwidth]{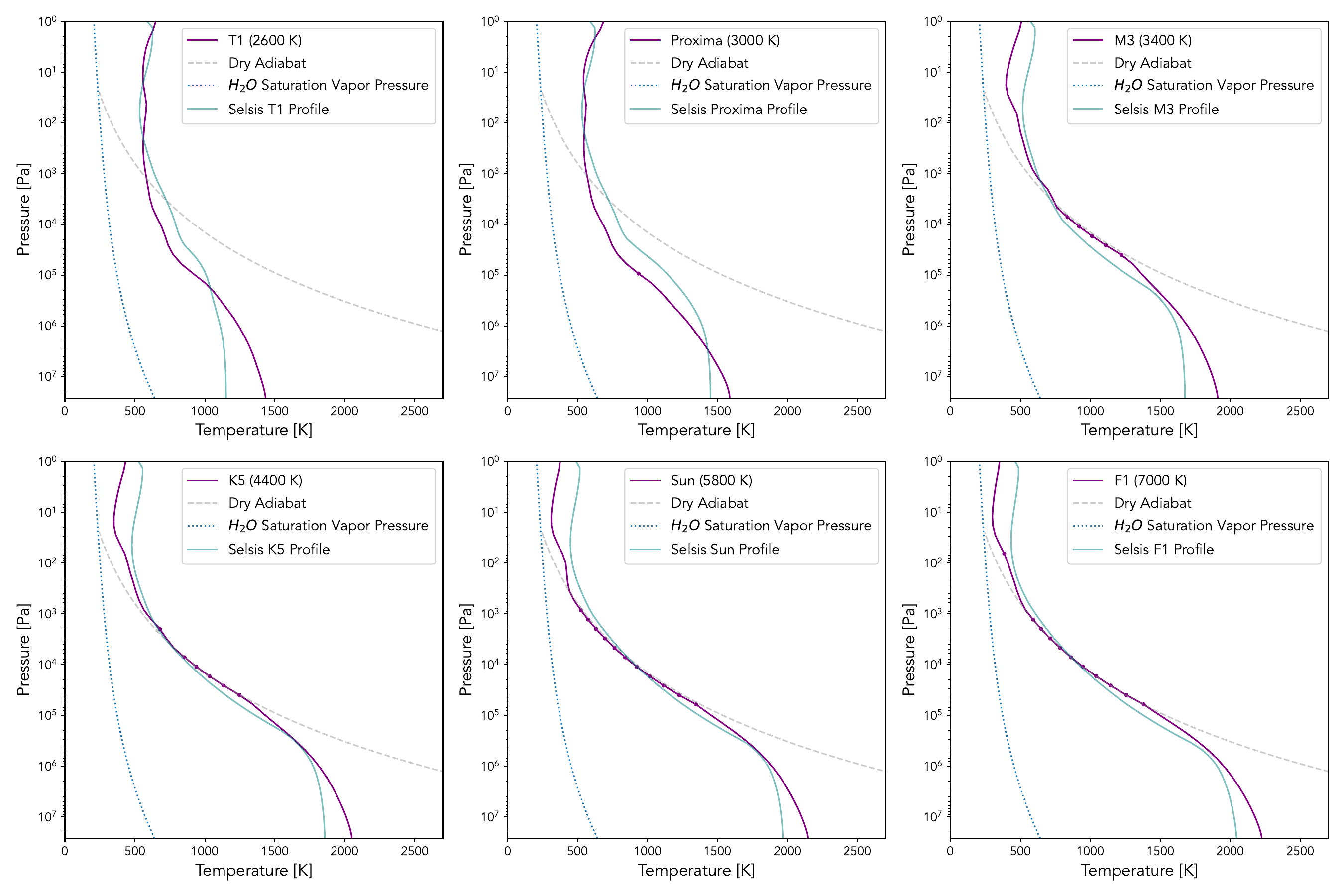}
    \caption{Temperature--pressure profiles for pure \ce{H2O} atmospheres irradiated by blackbody spectra at different stellar effective temperatures ($T_\mathrm{eff}$), all at OLR=$10^4$\,W\,m$^{-2}$. Dotted regions indicate convection. This figure isolates the effect of stellar temperature assuming an idealized Planck function for the incident flux.
    Results using both stellar spectra and blackbody spectra reveal a radiative zone near the surface, indicating that fully convective columns are not guaranteed in steam atmospheres.}
    \label{fig:Selsiscomparison}
\end{figure*}

Figure~\ref{fig:Selsiscomparison} compares our line-by-line (LBL) PCM-HiPT results with the correlated-$k$ simulations of \citet{Selsis2023cool}, both at an outgoing longwave radiation (OLR) of $10^4\,\mathrm{W\,m}^{-2}$. Our LBL model yields higher near-surface temperatures --- by up to a couple hundred kelvin ---underscoring how detailed absorption in the near-IR can shift where the radiative zone forms. Despite this offset, both approaches confirm that the lower atmosphere can be radiatively stable, detaching surface convection from mid-level convection.

\subsection{Spectral and Vertical Distribution of Radiative Fluxes}
\label{subsec:spectral_fluxes}

\begin{figure*}[ht!]
\centering
\includegraphics[width=\textwidth]{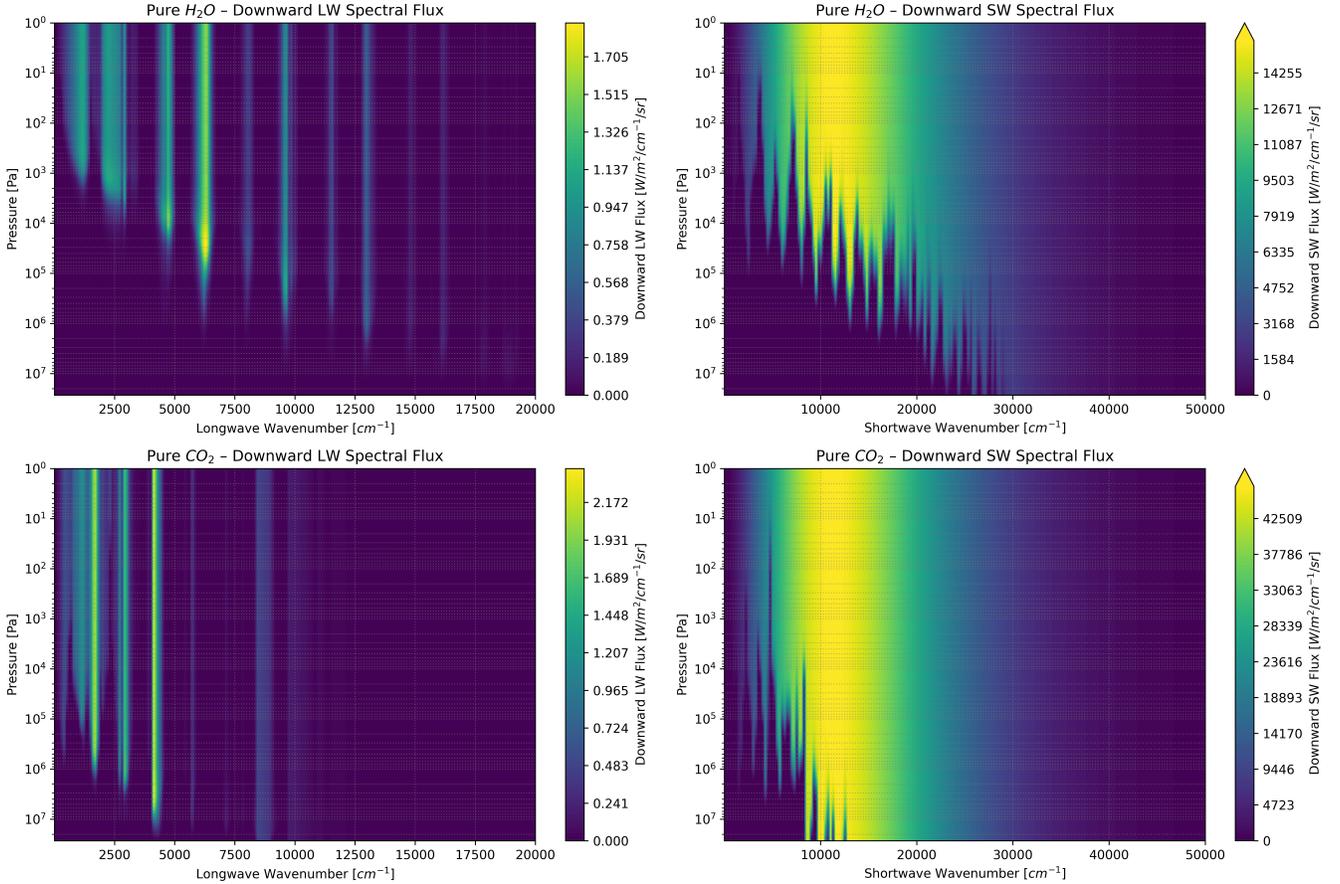}
\caption{Spectral fluxes vs.\ pressure in pure \ce{H2O} and \ce{CO2} atmospheres under a solar spectrum, at OLR=$10^4\,\mathrm{W\,m}^{-2}$. Top row: pure \ce{H2O} atmosphere. Bottom row: pure \ce{CO2} atmosphere. Left: net longwave (LW). Right: net shortwave (SW) flux.
Strong near-infrared absorption by \ce{H2O} creates a mid-atmospheric radiative layer that suppresses near-surface convection}

\label{fig:fluxmaps}
\end{figure*}

In Figure~\ref{fig:fluxmaps}, strong \ce{H2O} (top row) near-infrared absorption reduces the SW flux rapidly with height, creating a warm upper atmosphere and stabilizing lower layers against convection. In contrast, \ce{CO2} (bottom row) permits more stellar radiation to reach higher pressures, inhibiting deep heating and lowering surface temperatures. In both gases, LW absorption is prominent but does not fully offset the effects of SW absorption in shaping the temperature profile. Overall, these flux distributions demonstrate how upper-atmospheric shortwave absorption reduces near-surface convection and yields cooler surfaces than fully convective models would predict.

\subsection{Dependence on Stellar Type}
\label{subsec:stellar_type}

\begin{figure}[ht!]
    \centering
    \includegraphics[width=\columnwidth]{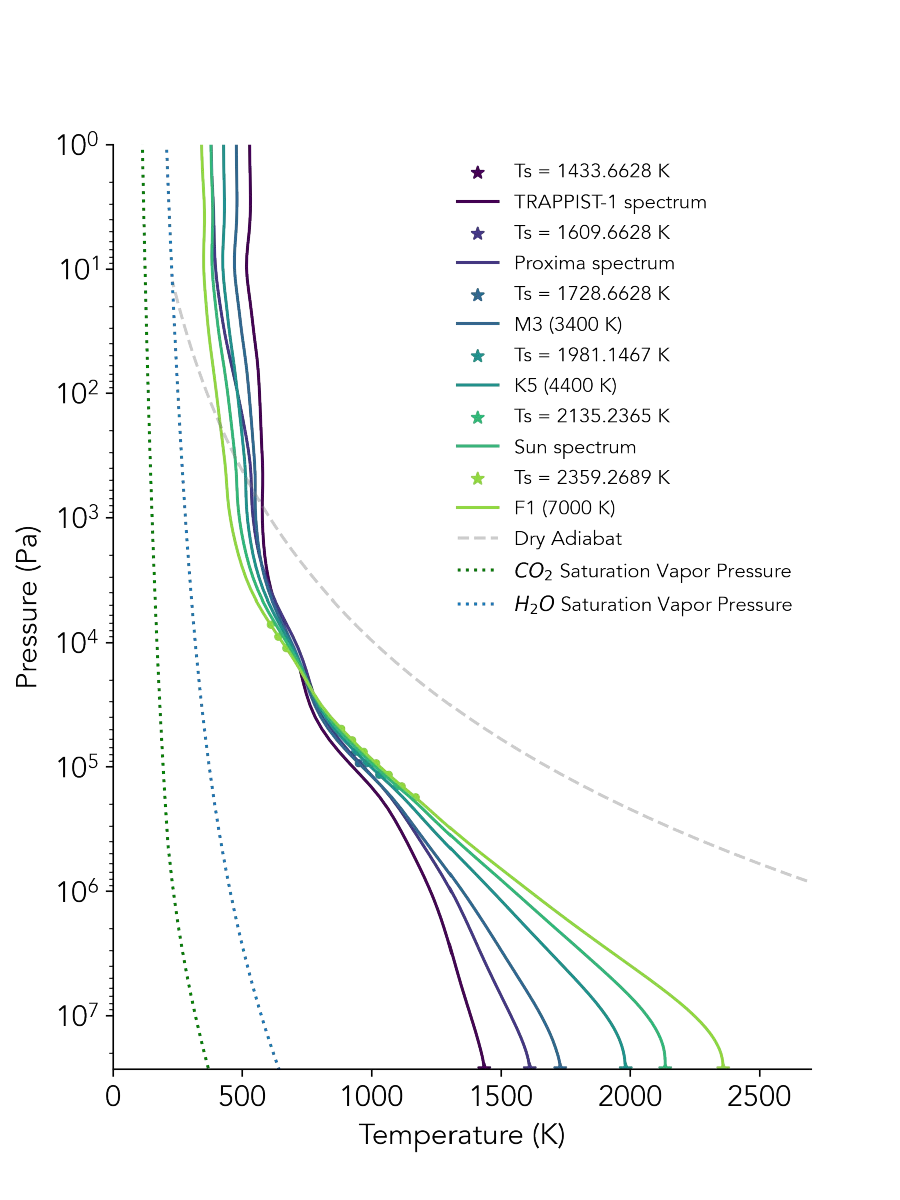}
    \caption{Temperature--pressure profiles for pure \ce{H2O} atmospheres under different stellar spectra, all at OLR=$10^4$\,W\,m$^{-2}$. Dotted regions indicate convection. Unlike Figure~\ref{fig:Selsiscomparison}, which assumes blackbody spectra for various stellar types, this figure uses observed stellar spectra for the Sun, TRAPPIST-1, and Proxima Centauri.
    More flux in \ce{H2O} near-IR bands (e.g., from hotter stars) raises the surface temperature by up to several hundred K, but a radiative zone still forms near the surface.}
    \label{fig:stellartypes}
\end{figure}

We next explore how the stellar spectrum alters the near-surface radiative zone. Figure~\ref{fig:stellartypes} shows temperature--pressure (T--p) profiles for pure \ce{H2O} atmospheres irradiated by stars of different effective temperatures, each converged to OLR=$10^4\,\mathrm{W\,m}^{-2}$.

Surface temperatures can differ by a few hundred kelvin, driven by how each spectrum overlaps with \ce{H2O} absorption bands. However, in all cases, a stable radiative layer appears below the mid-atmosphere convection, confirming that near-infrared opacity remains critical for preventing fully adiabatic profiles down to the ground.

Correlated-$k$ binning may underestimate strong near-IR \ce{H2O} absorption when flux peaks near 1--2\,$\mu$m. Our LBL code resolves thousands of pressure-broadened lines, depositing shortwave flux deeper and inhibiting convection. Additionally, \citet{Selsis2023cool} employs the ExoMol database, which differs from HITRAN and HITEMP in spectral line completeness, pressure broadening, and line wing treatments. These differences may lead to variations in absorption strengths, influencing the depth at which radiation is absorbed and subsequently impacting lower-atmosphere temperatures.

\subsection{Mixed \ce{H2O}-\ce{CO2} Atmospheres: Key Differences}
\label{subsec:mixed_comp}

\begin{figure*}[ht!]
    \centering
    \includegraphics[width=\textwidth]{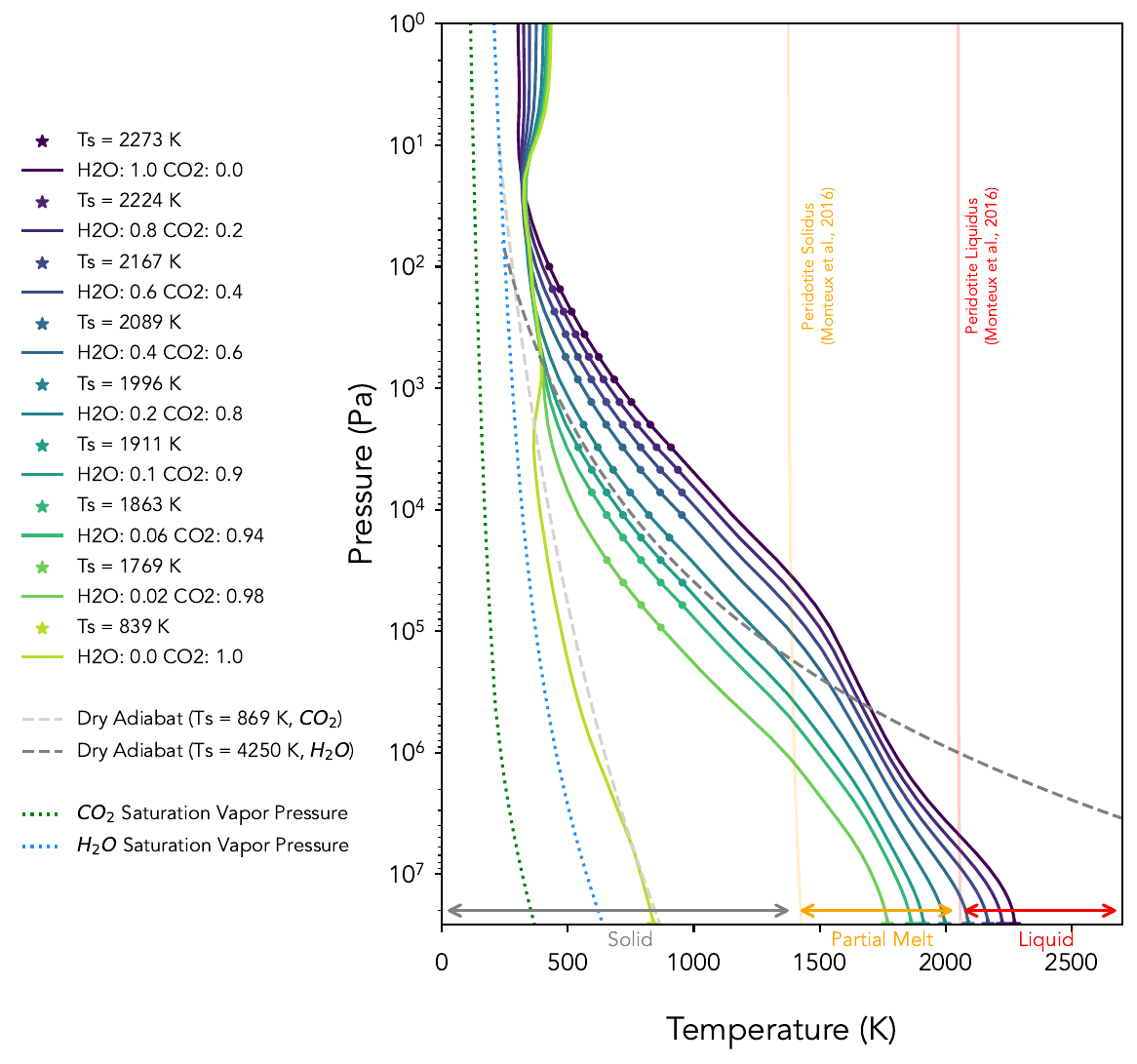}
    \caption{T--p profiles for various \ce{H2O}-\ce{CO2} mixtures, converged at ISR=$10^4\,\mathrm{W\,m}^{-2}$. Dotted regions indicate convection. Also shown are melt curves for peridotite solidus and liquidus \citep{Monteux2016}.} 
    Pure-\ce{CO2} atmospheres remain relatively cool near the surface, but adding even a few percent \ce{H2O} dramatically increases $T_s$ by reinstating opacity in the thermal infrared.
    \label{fig:atmosprofco2}
\end{figure*}

While modeling pure-steam envelopes helps isolate the climatic impact of \ce{H2O}, rocky exoplanets with magma oceans are expected to have abundant atmospheric \ce{CO2} under oxidizing or weakly reducing conditions \citep{sossi2020redox}. Figure~\ref{fig:atmosprofco2} compares T--p profiles for different \ce{H2O}:\ce{CO2} mixing ratios at the same ISR of $10^4\,\mathrm{W\,m}^{-2}$.
This corresponds to a planet at 0.37 AU from a Sun-like star. 

Pure \ce{CO2} yields a cooler surface $\sim\!1200$ and a more extended radiative region. In contrast, injecting even a few percent \ce{H2O} restores strong 1--3\,$\mu$m opacity, raising $T_s$ by hundreds of kelvin. Thus, small water abundances can ``tip" a \ce{CO2}-dominated atmosphere back into a steam-like regime with near-surface radiative inhibition and higher melt potential.

Figure~\ref{fig:atmosprofco2} also plots the peridotite solidus/liquidus \citep{Monteux2016}, indicating whether surface conditions permit partial or full melting of typical mantle rock. Under pure \ce{CO2}, the surface remains below the solidus, implying no sustained magma ocean. By contrast, 2--10\% \ce{H2O} can keep $T_s$ within the 1500--2000\,K range, sustaining partial melts for extended periods.

Hence, the competition between \ce{CO2} (which has narrower absorption bands at 2.7, 4.3, 15\,$\mu$m) and \ce{H2O} (with broad near-IR coverage) strongly modulates how much of the mantle might stay molten over geological timescales.

As a test, we also calculated surface temperature with 92 bar \ce{CO2}, 50 ppmv \ce{H2O} and a Venus-like ISR of 2600 W\,m$^{-2}$ and found $T_{surf}$ = 722 K (results not shown). This indicates that \ce{CO2} is not an effective greenhouse gas at high concentration.

\subsection{Trace Gases in \ce{CO2}-Dominated Atmospheres}
\label{subsec:trace_gases}

\begin{figure*}[ht!]
    \centering
    \includegraphics[width=\textwidth]{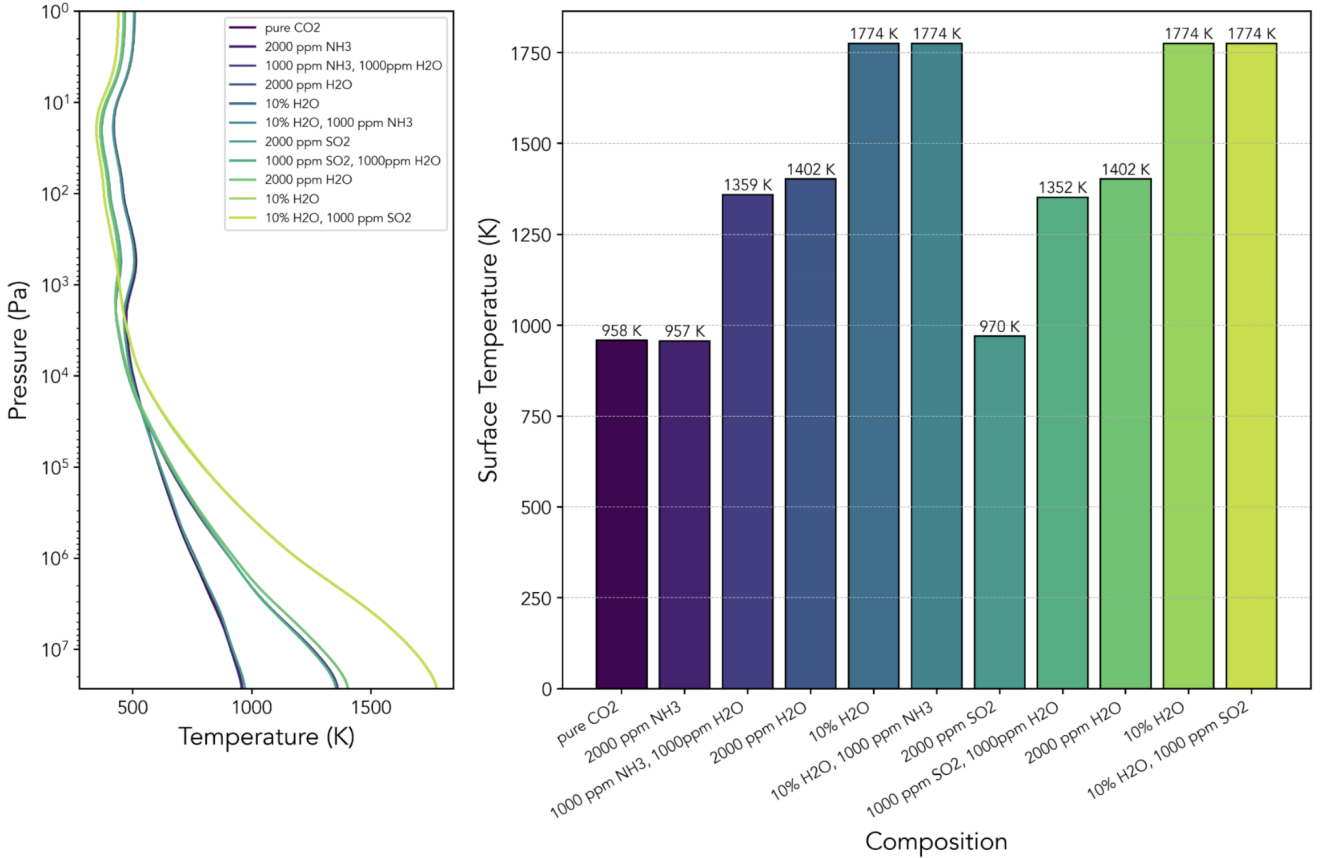}
    \caption{T--p profiles for trace \ce{SO2}, \ce{NH3} in \ce{CO2}-dominated atmospheres, all converged at OLR=$10^4\,\mathrm{W\,m}^{-2}$.} 
    Minor species matter primarily in compositions without significant water content; strong \ce{H2O} near-IR absorption otherwise overshadows their greenhouse effect.
    \label{fig:SO2_NH3}
\end{figure*}

Some volcanically active planets may release \ce{SO2} or \ce{NH3}, potentially altering the radiative balance. We test minor additions (1000--2000\,ppm) of these gases to a \ce{CO2}--rich mixture. Results (Figure~\ref{fig:SO2_NH3}) show that they can warm the near-surface region by up to a few hundred kelvin if \ce{H2O} is scarce, but become negligible once water vapor dominates.

This suggests that sulfur gases, ammonia, or other minor volatiles are most important for climates with limited water vapor. Once a few percent \ce{H2O} accumulates, water's broad absorption bands drive the formation of near-surface radiative layers.

\subsection{Outgoing Longwave Radiation (OLR) Spectra and Optical Depths}
\label{subsec:olr_spectra}

\begin{figure*}[ht!] 
\centering 
\includegraphics[width=\textwidth]{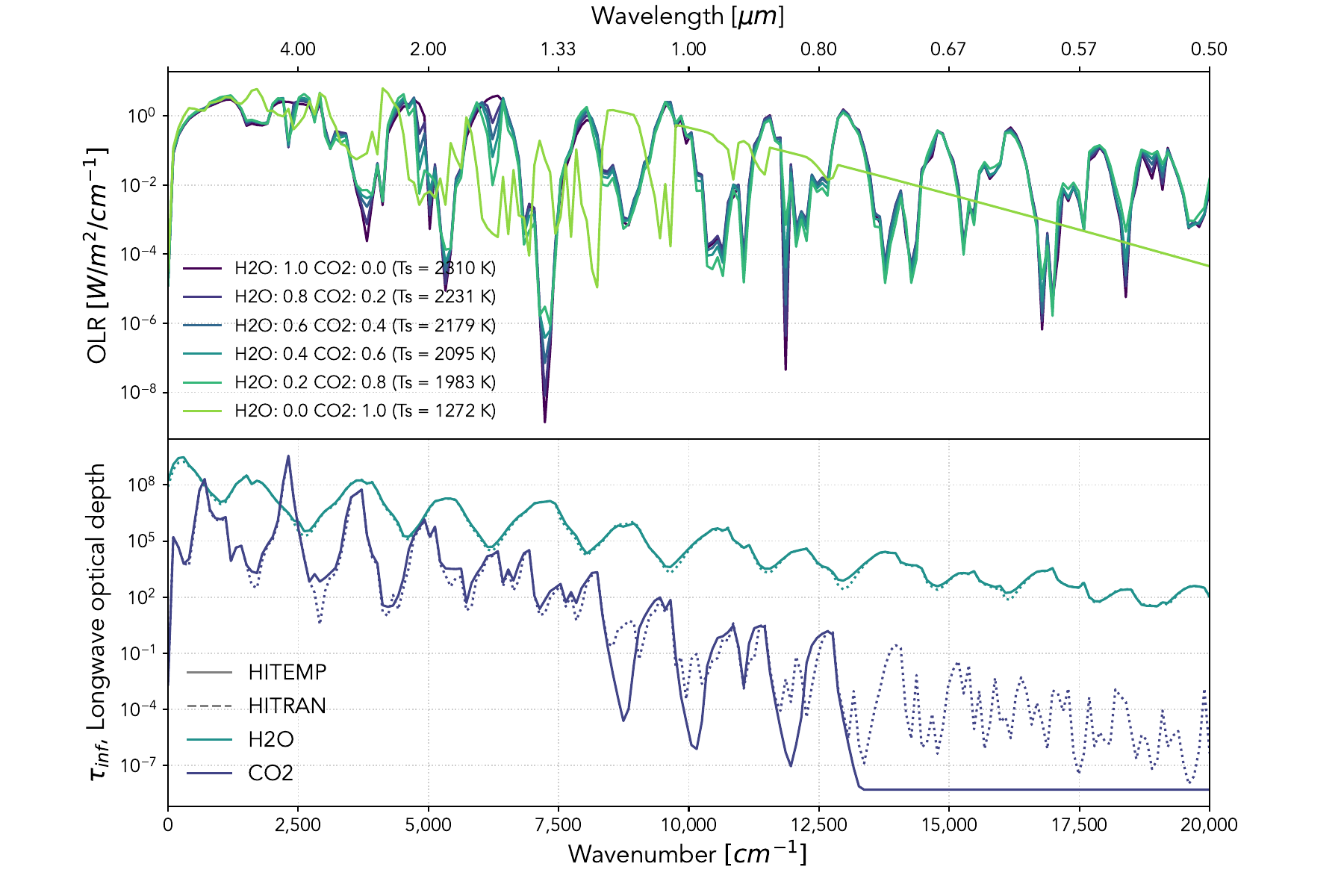} 
\caption{(Top) Outgoing longwave radiation (OLR) spectra for various \ce{H2O}-\ce{CO2} mixtures at OLR=$10^4\,\mathrm{W\,m}^{-2}$. (Bottom) Corresponding longwave (LW) optical depths, computed for the entire atmospheric column at the equilibrium temperature-pressure profiles from our model. Unlike the OLR, which represents emission at the top of the atmosphere, the optical depth values integrate absorption over the full column, showing how different spectral windows are affected by varying \ce{H2O} fractions.
Water vapor imposes broad absorption, limiting IR windows; small fractions of \ce{H2O} in \ce{CO2} quickly shift OLR toward steam-like behavior.}
\label{fig:OLR_tau}
\end{figure*}

To visualize how different compositions radiate heat, we plot OLR spectra for pure or mixed atmospheres at the same OLR=$10^4\,\mathrm{W\,m}^{-2}$. Figure~\ref{fig:OLR_tau} (top) shows that pure \ce{H2O} has broad emission features restricted by strong opacity, whereas pure \ce{CO2} has narrower bands, letting more IR escape, which lowers $T_s$. Injecting \ce{H2O} broadens near-IR absorption, elevating surface temperatures accordingly.

Optical depth comparisons (bottom of Fig.~\ref{fig:OLR_tau}) indicate minor differences between HITRAN and HITEMP data sets in these specific conditions, but neither removes the fundamental effect of strong near-IR absorption by \ce{H2O}. Overall, the key driver of near-surface heating is whether water vapor saturates the relevant shortwave bands.

\subsection{Summary of Key Trends}
In this section, we showed that:
\begin{enumerate}
    \item Pure-\ce{H2O} atmospheres form stable lower radiative zones even at high OLR, consistent with \citet{Selsis2023cool} but with slightly higher $T_s$ from our LBL approach.
    \item Pure-\ce{CO2} atmospheres remain cooler near the surface, often precluding magma oceans. Adding only a few percent \ce{H2O} elevates $T_s$ by a thousand kelvins.
    \item Trace gases (\ce{SO2}, \ce{NH3}) can matter for cases without \ce{H2O} but are overshadowed by broad \ce{H2O} absorption in steam-dominated scenarios.
    \item Surface temperatures exceeding the mantle solidus require moderate-to-high \ce{H2O} fractions, suggesting that outgassing history and redox state (favoring \ce{CO2} or \ce{H2O}) critically determine whether magma oceans persist.
\end{enumerate}

Taken together, these results emphasize that shortwave \ce{H2O} absorption remains a dominant factor shaping hot terrestrial atmospheres, with \ce{CO2} significantly cooler unless water is also present. We now turn to the broader implications for planetary evolution and interior-atmosphere coupling.
\section{Discussion}
\label{sec:discussion}

\subsection{Atmospheric Composition and Magma Ocean Cooling at High Temperatures}

Many highly irradiated terrestrial planets are predicted to host hot, dense atmospheres, which can arise when stellar flux pushes surface temperatures well beyond water's boiling point or through volcanic outgassing of greenhouse gases. Although classic models often label these states as ``post-runaway greenhouse," our work highlights that a range of near-surface temperatures can develop once large amounts of \ce{H2O} or \ce{CO2} accumulate. Notably, the formation of near-surface radiative layers is not confined to strictly post-runaway conditions but can occur whenever strong absorption in the near-infrared suppresses convection from the ground.

In \ce{CO2}-rich atmospheres, no convection is present, resulting in a more uniformly stratified thermal structure. Consequently, our results extend \citet{Selsis2023cool}---who showed that pure-steam atmospheres need not be fully convective---to mixtures of \ce{CO2} and \ce{H2O}, plus minor gases. Because \ce{CO2} is less soluble in magma than \ce{H2O}, many young terrestrial planets might outgas large quantities of \ce{CO2} \citep{Sossi2020}, creating cool ($<1500$\,K) surfaces unless sufficient water enters the gas phase. Even a small \ce{H2O} inventory can revert the lower column to a steam-like regime, however, raising $T_s$ by hundreds of kelvin.

\subsubsection{Melt Fraction and Interior Solidification}

\begin{figure*}[ht!]
    \centering
    \includegraphics[width=2.0\columnwidth]{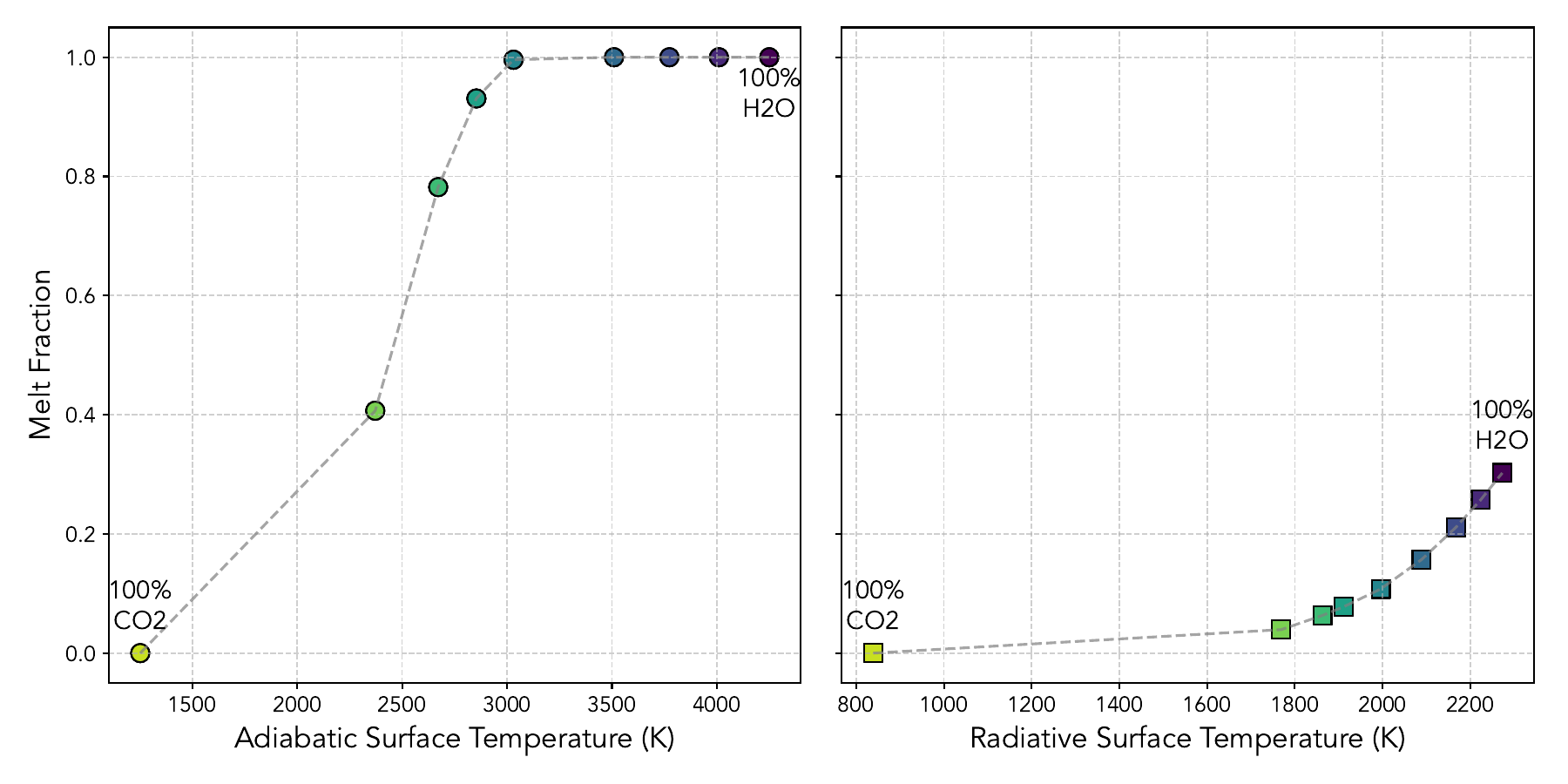}
    \caption{Melt fraction as a function of adiabatic and radiative surface temperatures, illustrating the transition from partially molten to fully solidified states for ISR = $10^4\,\mathrm{W\,m}^{-2}$. Surface temperatures correspond to those of each curve in Figure \ref{fig:atmosprofco2}. High surface temperatures sustain widespread melt, while cooling leads to rapid solidification. The melt fraction evolution plays a critical role in volatile sequestration and outgassing dynamics. The labeled endpoints correspond to pure \ce{CO2} (100\% \ce{CO2}, 0\% \ce{H2O}) on the left and pure \ce{H2O} (0\% \ce{CO2}, 100\% \ce{H2O}) on the right, with intermediate points representing mixed compositions. The compositional fractions for each point, from left to right, are: 98\% \ce{CO2} / 2\% \ce{H2O}, 94\% \ce{CO2} / 6\% \ce{H2O}, 90\% \ce{CO2} / 10\% \ce{H2O}, 80\% \ce{CO2} / 20\% \ce{H2O}, 60\% \ce{CO2} / 40\% \ce{H2O}, 40\% \ce{CO2} / 60\% \ce{H2O}, 20\% \ce{CO2} / 80\% \ce{H2O}, and 0\% \ce{CO2} / 100\% \ce{H2O}, indicating a progressive shift in atmospheric composition across the sequence.}
    \label{fig:melt_fraction}
\end{figure*}

To quantify how surface temperature controls melt retention, Figure~\ref{fig:melt_fraction} illustrates the melt fraction across different compositions, from pure \ce{CO2} to pure \ce{H2O}. When $T_s$ exceeds $\sim$1500--1700\,K, extensive melting can persist, affecting how volatiles such as \ce{CO2} and \ce{H2O} partition between the magma and atmosphere. As the planet cools, solidification may occur abruptly or in stages, influencing outgassing pathways and volatile sequestration in the crust \citep{Kite2009, Hu2014}. 
Once the mantle fully solidifies, geochemical and atmospheric processes---such as weathering, carbonate-silicate cycling, and long-term atmospheric escape---reshape atmospheric composition. If surface temperatures drop below the critical point of \ce{H2O}, condensation and silicate weathering may draw down large portions of \ce{CO2} and \ce{H2O}, leading to a drier atmosphere. Additionally, interactions between the crust and residual volatiles can buffer atmospheric evolution, influencing the rate of \ce{CO2} sequestration into the interior \citep{Kite2009, Hu2014}. Over time, continued hydrogen loss via atmospheric escape could further oxidize the atmosphere, favoring the buildup of \ce{O2}-rich or \ce{CO}-dominated compositions \citep{Wordsworth2018, Sossi2020}.

\subsection{Climate Hysteresis and the Role of Initial Conditions}

An interesting implication of our work is the potential for climate hysteresis in hot terrestrial atmospheres. If a planet begins with a \ce{CO2}-rich inventory without \ce{H2O}, strong radiative cooling might prevent surface temperatures from reaching the \ce{H2O} outgassing threshold. Such a planet could potentially remain locked in a stable, cool state unless a large perturbation---e.g., intense volcanism---injects significant water vapor. Conversely, if water is abundant early on, near-IR absorption can keep $T_s$ high, promoting further \ce{H2O} outgassing and possibly extending the magma ocean phase. These scenarios underscore how a planet's atmospheric history can dictate its long-term thermal and chemical evolution.

\subsection{Volatile Partitioning and Redox Evolution}
\label{sec:mo_solidification}

Near-surface radiative inhibition also affects the fraction of magma in the upper mantle, altering how \ce{H2O} and \ce{CO2} partition between the interior and atmosphere. If \ce{H2O} tends to remain dissolved while \ce{CO2} outgasses more readily \citep{Sossi2022}, then despite a large initial water budget, the final atmospheric composition could be \ce{CO2}-dominated, especially if hydrogen escape oxidizes the surface over time \citep{hamano2013emergence, Wordsworth2018}.

At intermediate stages, partial melting and selective degassing can yield a mixed \ce{CO2}-\ce{H2O} atmosphere, still prone to radiative inhibition near the surface. Once the mantle fully solidifies, residual \ce{H2O} remains locked in the interior, influencing potential rehydration events if tectonic processes become active. More comprehensive, time-resolved simulations that couple interior cooling, volcanism, and atmospheric radiative transfer [e.g., \citet{lebrun2013, Turbet2021}] will be essential for quantifying these feedbacks.

Beyond the \ce{CO2}-\ce{H2O} mixtures considered here, additional atmospheric compositions could emerge depending on redox state and atmospheric escape processes. Hydrogen loss from a primordial steam atmosphere, for example, could leave behind an \ce{O2}-rich residual atmosphere, influencing subsequent oxidation pathways \citep{Sossi2020}. In high-temperature regimes, photodissociation of \ce{H2O} can enhance \ce{CO} production \citep{Hu2014}, while sufficiently high outgassing temperatures may allow \ce{H2} to persist, particularly on planets with substantial \ce{H2}-retaining reservoirs. The strong collision-induced absorption (CIA) of \ce{H2} in dense environments could significantly modify the radiative balance \citep{Borysow2002}. While we do not explore \ce{H2}- or \ce{CO}-dominated atmospheres in this study, they remain relevant for highly reduced conditions or planets undergoing atmospheric escape. Future studies incorporating \ce{H2} and reduced carbon species will refine our understanding of thermal structure in extreme planetary environments.

\subsubsection{Impact of Trace Gases and Redox State}
\label{sec:trace_gases_redox}

We examine the role of trace species (\ce{SO2}, \ce{NH3}) in \ce{CO2}-dominated atmospheres, focusing on their ability to modify surface temperatures under varying \ce{H2O} concentrations. Our results confirm that water vapor remains the dominant control on atmospheric thermal structure, while the influence of trace gases depends strongly on baseline \ce{H2O} abundance.

Figure~\ref{fig:SO2_NH3} illustrates how surface temperature varies across different compositions, explicitly distinguishing cases with 1000 ppm and 2000 ppm of each species. The key takeaway is that adding \ce{NH3} or \ce{SO2} alone has a more pronounced warming effect when \ce{H2O} is scarce, while at higher \ce{H2O} abundances, their influence becomes negligible.

In \ce{CO2}-dominated scenarios with minimal \ce{H2O}, \ce{NH3} and \ce{SO2} contribute additional greenhouse heating, but once \ce{H2O} reaches a few percent by volume, its broad near-IR opacity overshadows any trace-gas effects.

For the \ce{NH3} runs, the \ce{CO2} + 1000~ppm \ce{NH3} + \ce{H2O} case has a higher surface temperature than the \ce{CO2} + 2000~ppm \ce{NH3} case, because of the strong greenhouse effect of \ce{H2O}.

Real exoplanets could outgas other sulfur species (e.g., H\(_2\)S, OCS) or CO, depending on redox state and volcanic flux \citep{Hu2014,Gao2017}. Some molecules have strong absorption or form aerosols, potentially altering the radiative budget. Photochemical and aerosol processes remain beyond this study's scope but may introduce new feedbacks (e.g., haze formation) that either enhance or mitigate near-surface radiative inhibition.

\subsection{Compositional Implications and Observational Prospects}
\label{sec:observations}

Recent work has emphasized that the high solubility of \ce{H2O} relative to \ce{CO2} generally limits the formation of steam atmospheres, except under moderately oxidizing conditions with high H/C ratios \citep{Sossi2023}. However, if a planet hosts only a limited or transient magma ocean, there is correspondingly less melt available to dissolve \ce{H2O}. This reduced capacity for water incorporation into the melt means that the atmospheric water budget could be significantly affected compared to scenarios with extensive magma oceans. Similar effects may occur for other soluble species.

Observationally, this suggests that compositional differences---such as the atmospheric abundance of \ce{H2O}---could be a powerful way to test for deep atmospheric radiative layers indirectly. High-resolution spectroscopic measurements targeting key volatile species may therefore offer critical insights into the thermal evolution and interior dynamics of hot terrestrial planets. Future coupled atmosphere-interior modeling studies and multi-wavelength observations will be essential to test these predictions.

\subsection{Modeling Limitations} 
\label{sec:limitations}

Our upper spectral cutoff of $20{,}000$\,cm$^{-1}$ (0.5\,\(\mu\)m) and lack of UV photochemistry are key caveats. Shortwave absorption above 0.5\,\(\mu\)m, especially in the UV, could modify upper-atmosphere heating or drive additional chemistry \citep[e.g.,][]{Tian2010}. Extending line-by-line calculations beyond $20{,}000$\,cm$^{-1}$ with improved high-T continuum data \citep{Kitzmann2017} is an important future step for refining near-surface conditions, especially around hotter stars.

Moreover, we assume steady-state radiative equilibrium at a fixed OLR. Real planets might have significant internal fluxes from cooling mantles or variability in stellar flux, driving the system out of equilibrium at certain epochs. Addressing such transient phases requires fully time-dependent modeling, which lies outside this work's scope.

To fully capture these effects, future work should adopt time-dependent, interior-atmosphere coupled models that track the interplay between radiative inhibition, mantle cooling, and episodic outgassing events. A time-integrating approach would clarify whether radiative layers dynamically evolve as the planet loses heat, or if shortwave absorption maintains them long enough to significantly alter solidification timescales.

\section{Conclusion}
\label{sec:conclusion}

We have developed and applied PCM-HiPT, a one-dimensional line-by-line radiative-convective model designed for dense, hot exoplanet atmospheres, with a particular focus on \ce{CO2}-\ce{H2O} mixtures. Our key conclusions are:

\begin{enumerate}
    \item Near-surface radiative zones, rather than full convection, may be common in high-temperature \ce{H2O}-rich atmospheres. Shortwave absorption by \ce{H2O} can suppress convection from the ground up, reducing $T_s$ by up to ~2000 K compared to fully adiabatic models.
    \item \ce{CO2}-dominated atmospheres tend to be cooler near the surface, but even minor \ce{H2O} fractions restore strong near-IR absorption. For example, introducing just 2\% \ce{H2O} to an otherwise \ce{CO2}-dominated atmosphere increases $T_s$ by up to 1000 K, shifting the system back toward a structure characteristic of \ce{H2O}--dominated atmospheres.
    \item The likelihood of long-lived magma oceans is strongly tied to atmospheric composition. Pure \ce{CO2} worlds tend to solidify quickly, whereas \ce{H2O}-rich atmospheres can sustain molten surfaces for extended periods.
    \item Planets may experience climate hysteresis, where initial volatile inventory dictates whether they stabilize in a hot steam-rich state or a cooler \ce{CO2}-dominated regime.
\end{enumerate}

These findings have several implications for terrestrial planet evolution, especially during or after ``runaway greenhouse" phases. Our results support the conclusion of \citet{Selsis2023cool} that near-surface radiative layers in \ce{H2O}-rich atmospheres will cause magma oceans to be less extensive and shorter-lived than earlier full-convection models suggested. 

Our work extends \citet{Selsis2023cool} by exploring mixed composition atmospheres. When both \ce{CO2} and \ce{H2O} are present, even minor \ce{H2O} additions cause rapid increases in surface temperature. Trace gases like \ce{SO2} or \ce{NH3} can cause additional warming but matter most in drier atmospheres.
Given the diversity of possible atmospheric chemistries, it is crucial to explore not only \ce{H2O}-\ce{CO2} mixtures but also atmospheres dominated by reducing gases (\ce{H2}, \ce{CH4}, \ce{CO}) or sulfur-rich conditions (\ce{SO2}), which have distinct radiative and photochemical properties \citep{Hu2014, Gao2017, Wordsworth2018}. While our study focuses on \ce{H2O}-\ce{CO2} atmospheres, reduced compositions with significant \ce{H2} or \ce{CH4} could exhibit enhanced collision-induced absorption (CIA) \citep{Abel2011}, while sulfur-rich atmospheres might experience strong aerosol-driven cooling. Additional species such as \ce{H2S}, and \ce{OCS} could further influence radiative balance by modifying atmospheric opacity \citep{Gao2017}.

Despite its advantages, PCM-HiPT faces limitations, especially the upper spectral limit of $20{,}000$\,cm$^{-1}$. Extending line-by-line calculations to higher wavenumbers and incorporating improved high-T continuum data will enhance accuracy for very hot planets orbiting bright stars. Time-dependent coupling between planetary interiors and atmospheres would also be useful for simulating long-term evolution and climate responses to stochastic events.

While the atmospheres explored in this study are too hot to be habitable, our findings may have implications for cooler planets, including at the runaway greenhouse transition. Future work should explore a wider range of stellar fluxes and volatile inventories to assess the importance of radiative effects on such planets.

Looking ahead, rocky exoplanet transit spectra could potentially test these predictions by probing key near-IR features of \ce{CO2}- versus \ \ce{H2O}-dominated atmospheres. The large decreases in surface temperature that result from deep radiative layers will also impact atmosphere-interior volatile exchange, which should have observable consequences. Future work coupling our model to chemical and interior models will allow these effects to be investigated in full.

\section{Code Availability}
The PCM-HiPT (Planetary Climate Model for High Pressures and Temperatures) source code is available at Zenodo (\href{https://doi.org/10.5281/zenodo.15278168}
{https://doi.org/10.5281/zenodo.15278168}).

\section{Acknowledgements}

The authors acknowledge funding from NSF CAREER award AST-1847120, NASA VPL award UWSC10439 and NSF award AGS-2210757. We also thank the anonymous reviewers for their insightful comments and suggestions. 

\bibliography{main}{}
\bibliographystyle{aasjournal}

\section*{Appendix}
\label{appendix}

\subsection{Hybrid Pressure Array: Derivation and Implementation}
\label{appendix:hybrid_array}

In dense, hot atmospheres, resolving the lower layers accurately (where optical depths can be extremely large) is crucial for capturing near-surface radiative processes. To avoid an excessive number of vertical layers, we employ a hybrid pressure array that blends linear spacing near the surface with logarithmic spacing aloft, ensuring sufficient resolution where it matters most.

We construct two arrays: a logarithmic array, \(P_{\text{log}}\), and a linear array, \(P_{\text{linear}}\). The final pressure distribution, \(P_{\text{hybrid}}\), smoothly transitions between them via a weighting function \(W(x)\).

\subsubsection{Logarithmic vs.\ Linear Arrays}

The logarithmic array can be generated using:
\begin{equation}
P_{\text{log}}(x) \;=\; 
10^{\Bigl(\log_{10}\bigl(P_{\text{min}}\bigr) 
\;+\; 
\frac{x\,\bigl[\log_{10}(P_{\text{max}}) - \log_{10}(P_{\text{min}})\bigr]}{N-1}\Bigr)},
\end{equation}
where \(x\) is the index (0 to \(N-1\)), and \(P_{\text{min}}\), \(P_{\text{max}}\) define the array bounds in pressure space.

The linear array can be generated as:
\begin{equation}
P_{\text{linear}}(x) 
\;=\;
P_{\text{min}}
\;+\;
\frac{x}{N}\,\bigl(P_{\text{max}} - P_{\text{min}}\bigr).
\end{equation}

\subsubsection{Combining the Arrays}

To produce the final hybrid array, \(P_{\text{hybrid}}(x)\), we interpolate between \(P_{\text{log}}(x)\) and \(P_{\text{linear}}(x)\) using a smooth weighting function \(W(x)\):
\begin{equation}
P_{\text{hybrid}}(x)
\;=\;
W(x)\,P_{\text{log}}(x)
\;+\;
\bigl[\,1 - W(x)\bigr]\,P_{\text{linear}}(x).
\end{equation}

We define
\begin{equation}
W(x) 
\;=\;
\frac12\Bigl(
1 + \tanh\!\Bigl(\frac{x - x_0}{\sigma}\Bigr)
\Bigr),
\end{equation}
where \(x_0\) is the center index of the transition and \(\sigma\) sets its width.

\subsubsection{Newton's Method for the Transition Point}
In some applications, one might wish to specify the exact pressure at which to switch from linear to log spacing. To find the index \(t\) where \(P_{\text{hybrid}}(t) = P^{*}\) (the desired transition pressure), we use Newton's method.

Hence, we solve:
\begin{equation}
W(t)\,P_{\text{log}}(t)
\;+\;
\bigl[\,1 - W(t)\bigr]\,P_{\text{linear}}(t)
\;=\;
P^{*}.
\end{equation}
We denote this function \(f(t)=P_{\text{hybrid}}(t)-P^{*}\), compute its derivative \(f'(t)\), and iteratively update \(t\leftarrow t - f(t)/f'(t)\). The expansions of \(f(t)\) and \(f'(t)\) involve chain-rule derivatives of the \(\tanh\)-based weighting function and the log/linear arrays.

\begin{equation}
\small
f(t) \;=\;
W(t)\,10^{\Bigl(\log_{10}(p_{\min}) + t\cdot\frac{\log_{10}(p_{\max}) - \log_{10}(p_{\min})}{N-1}\Bigr)}
\;+\;
\bigl[1 - W(t)\bigr]\,
\bigl(p_{\min} + t\,\tfrac{p_{\max}-p_{\min}}{N}\bigr)
\;-\;
P^{*}.
\end{equation}
\normalsize

After convergence, \(t\) specifies the array index (or fractional index) at which we place the transition. The resulting hybrid array accommodates high resolution near the surface while maintaining efficiency at lower pressures.

\begin{figure}[h!]
    \centering
    \includegraphics[scale=0.6]{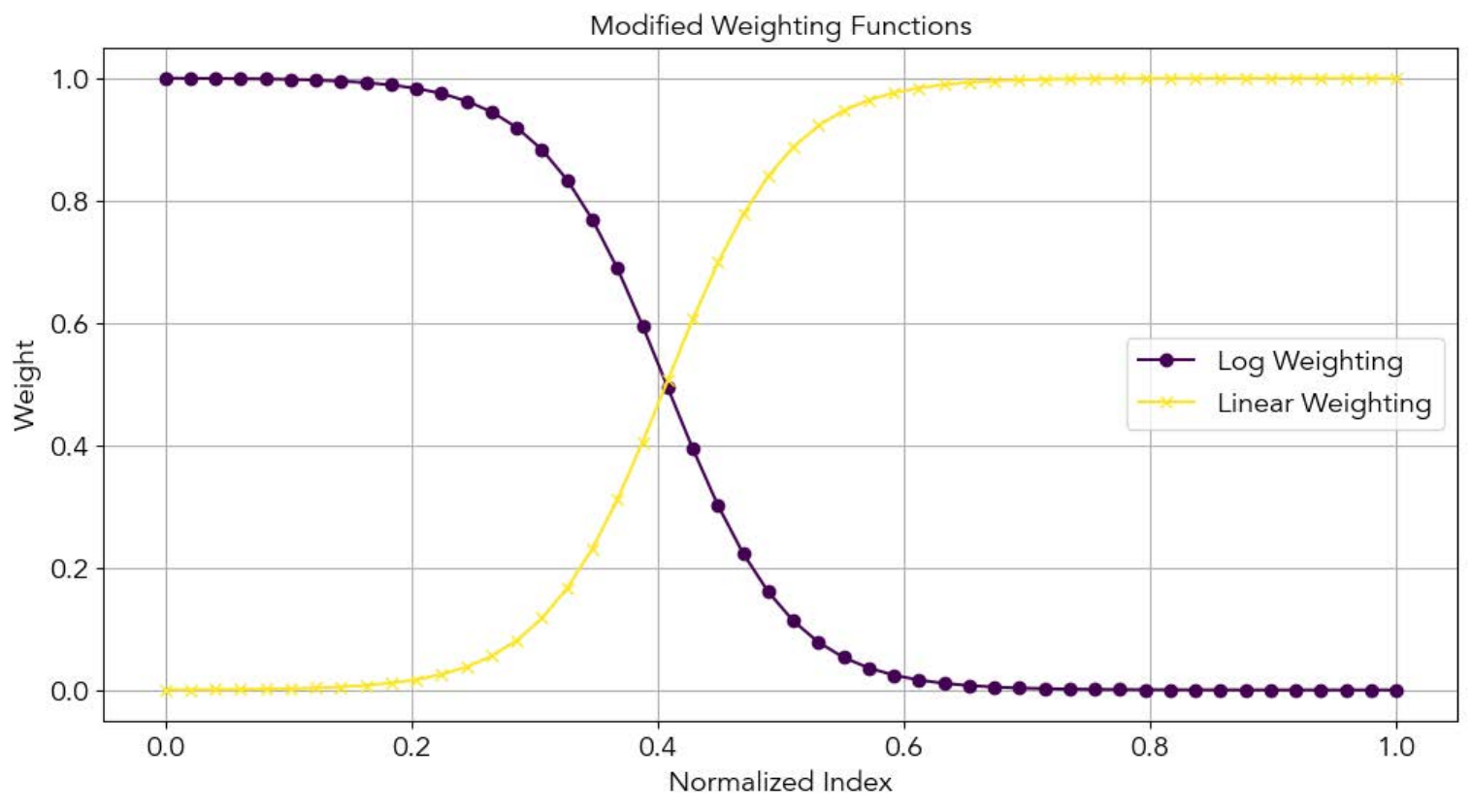}
    \caption{An example of the weighting functions used in the hybrid scheme.} The log weight (purple) and linear weight (yellow) vary smoothly with index \(x\), controlled by \(\sigma\) and \(x_0\).
    \label{fig:WeightingFunctions}
\end{figure}

\begin{figure}[h!]
    \centering
    \includegraphics[scale=0.5]{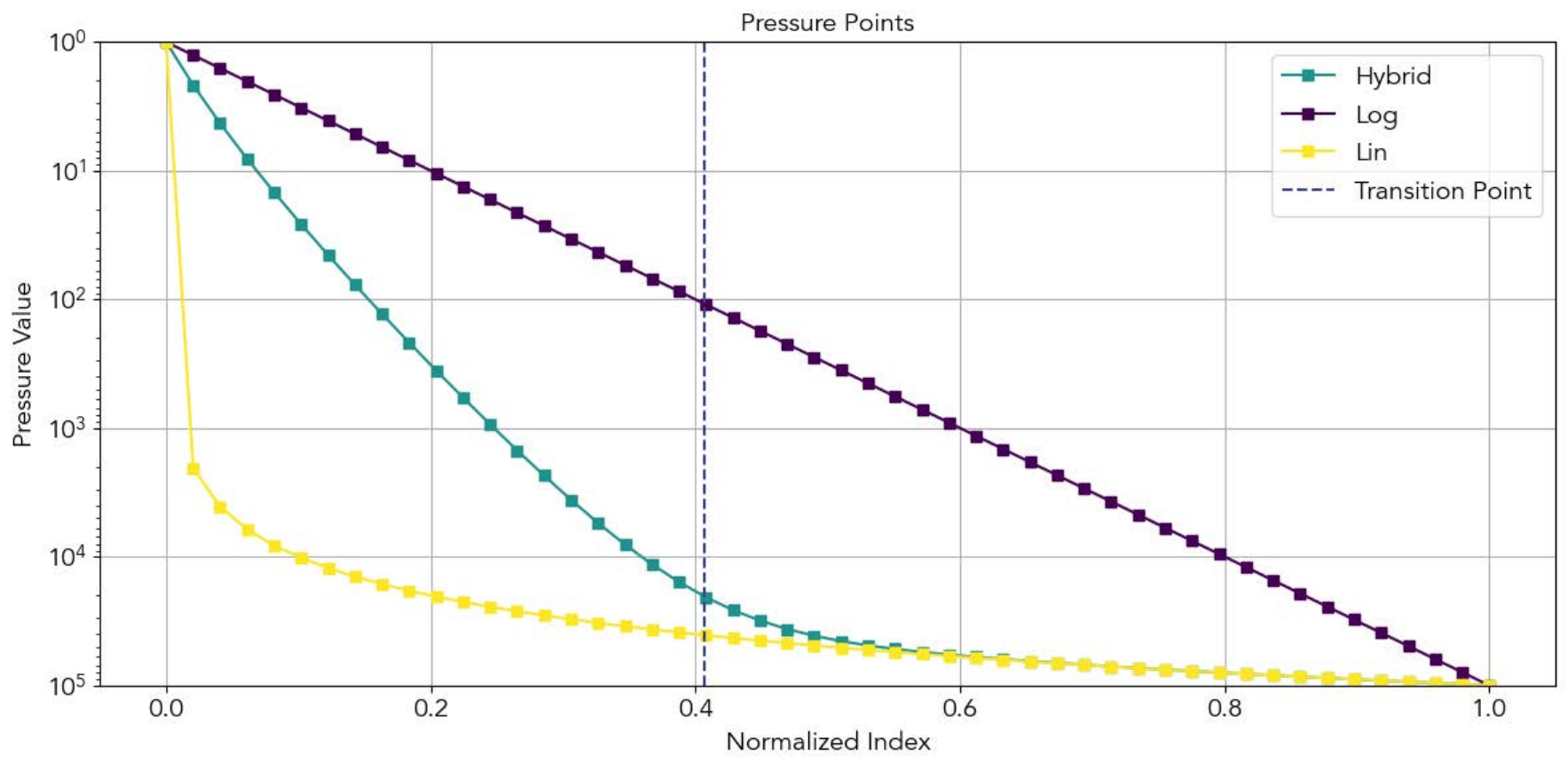}
    \caption{Pressure values generated using the hybrid approach. We linearly space levels in the dense lower atmosphere, then smoothly transition to log spacing aloft. The dashed vertical line shows the transition point.}
    \label{fig:HybridPressureArray}
\end{figure}

In summary, the hybrid approach ensures refined vertical resolution in optically thick lower layers without an unwieldy number of layers at high altitudes. This technique is widely applicable to 1-D radiative-convective modeling of exoplanets with high surface pressures.

\subsection{Comparison of HITEMP and HITRAN Spectroscopic Data}
\label{appendix:hitemp_hitran}

To assess how spectral data choices affect our atmospheric calculations, we compare results using the HITRAN2020 \citep{Gordon2022} and HITEMP2010 \citep{rothman2010hitemp} databases. While HITRAN is designed for lower temperatures and typically used in Earth-based applications, HITEMP includes high-temperature spectral lines relevant for exoplanet atmospheres.

\begin{figure}[ht!]
\centering
\includegraphics[width=0.7\columnwidth]{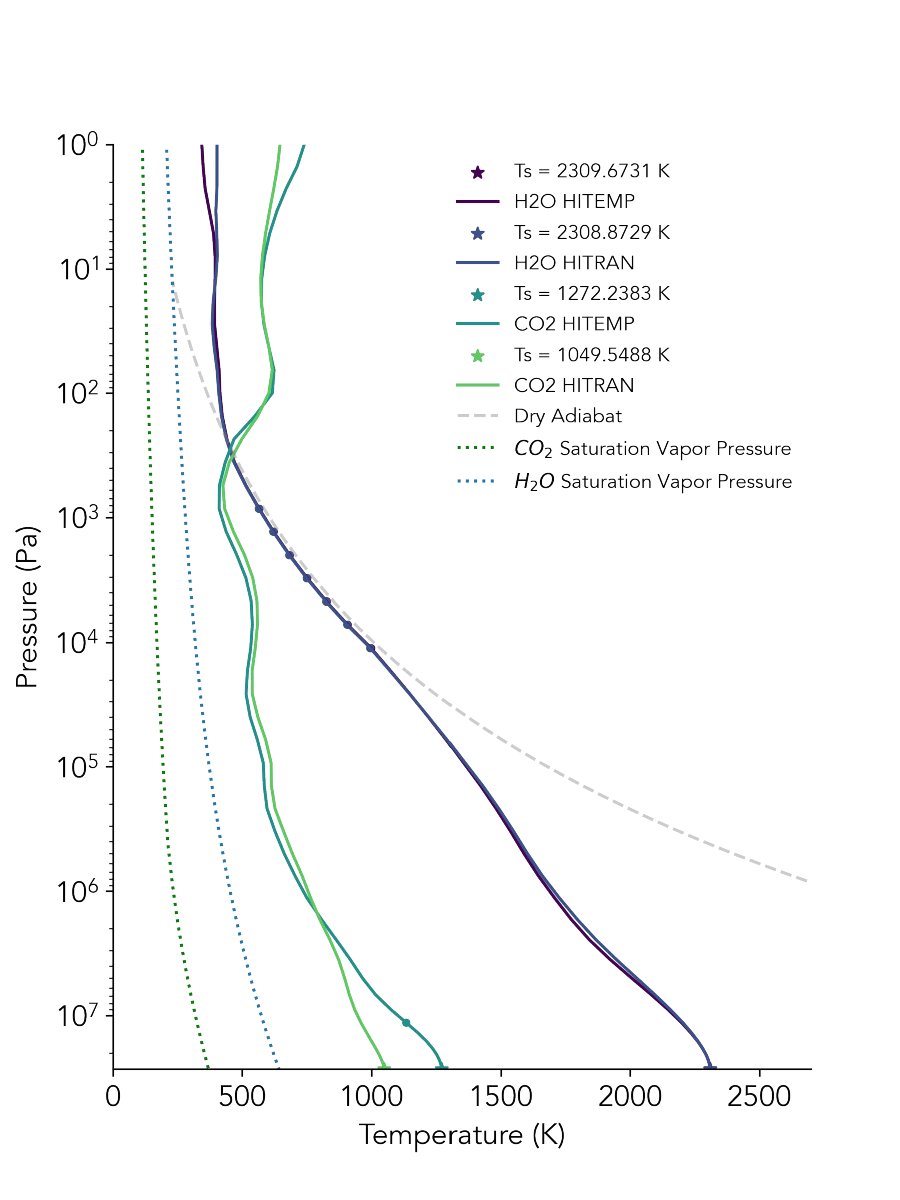}
\caption{Comparison of atmospheric profiles generated using HITRAN2020 and HITEMP2010 databases for an OLR of 10,000 W~m$^{-2}$ under a solar spectrum. Dotted regions indicate convection.
Differences between HITRAN and HITEMP are modest in our regime, suggesting HITRAN is sufficient for this study.}
\label{fig:hitran_hitemp_comparison}
\end{figure}

As shown in Fig.~\ref{fig:hitran_hitemp_comparison}, temperature differences between the two databases are minimal in the high-pressure, high-temperature regime we explore. This suggests that HITRAN provides sufficiently accurate results for our modeling, though more detailed high-temperature laboratory measurements could refine future calculations.

\subsection{Sensitivity of Atmospheric Structure to Heat Capacity ($c_p$)}
\label{appendix:cp_sensitivity}

The specific heat capacity ($c_p$) affects the adiabatic lapse rate and, in turn, the convective temperature structure. To test its influence, we vary $c_p$ between 2000 and 3500 J\,kg$^{-1}$\,K$^{-1}$ and compare the resulting T--p profiles.

\begin{figure}[ht!]
    \centering
    \includegraphics[width=0.7\columnwidth]{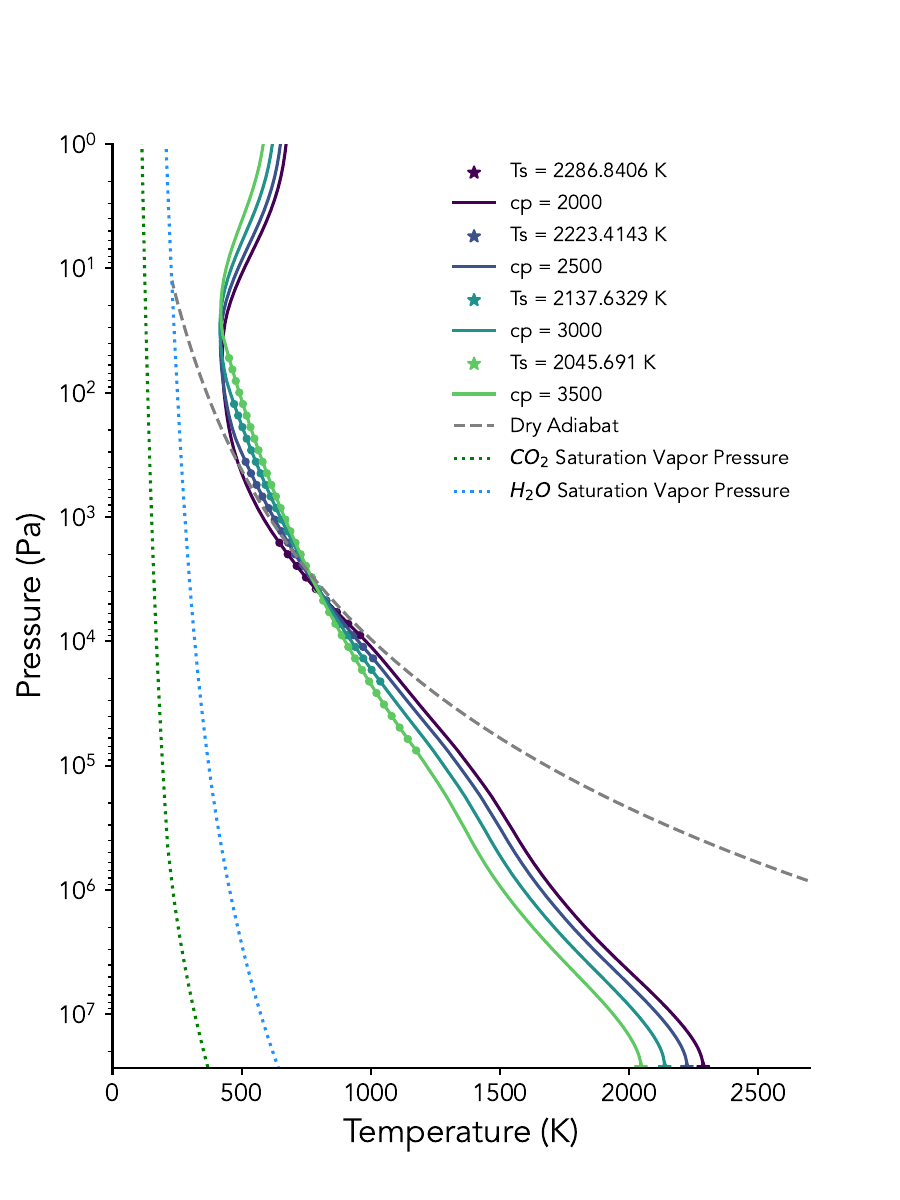}
    \caption{Sensitivity of atmospheric T--p profiles to constant heat capacity values. Dotted regions indicate convection. While the absolute temperatures shift slightly, the broad structure of radiative inhibition near the surface remains.}
    \label{fig:cp_sensitivity_example}
\end{figure}

As seen in Fig.~\ref{fig:cp_sensitivity_example}, variations in $c_p$ shift the temperature profile by tens of kelvin but do not alter the presence of near-surface radiative layers or detached convection zones. This confirms that using a constant $c_p$ (2085 J\,kg$^{-1}$\,K$^{-1}$) is a reasonable approximation for our study.

While more detailed treatments could incorporate $c_p(T, P)$ variations, our results indicate that radiative inhibition due to shortwave \ce{H2O} absorption is robust to such parameter changes.

\end{document}